\documentclass[epjc3]{svjour3} 

\usepackage[margin=2.5cm]{geometry}

\usepackage{amsmath,amssymb}

\usepackage[dvipdfmx]{graphicx} 
\usepackage{wrapfig}

\usepackage[xcolor,framemethod=tikz]{mdframed}

\newcommand{\Tph}{T_{\mathrm{ph}}}
\newcommand{\mcTsallisS}{S_{\mathrm{T}q}^{\mathrm{MC}}}
\newcommand{\mcRenyiS}{S_{\mathrm{R}q}^{\mathrm{MC}}}
\newcommand{\caTsallisS}{S_{\mathrm{T}q}^{\mathrm{C}}}
\newcommand{\caRenyiS}{S_{\mathrm{R}q}^{\mathrm{C}}}
\newcommand{\mcT}{T^{\mathrm{MC}}}
\newcommand{\mcTph}{T_{\mathrm{ph}}^{\mathrm{MC}}}
\newcommand{\cabeta}{\beta^{\mathrm{C}}}
\newcommand{\caT}{T^{\mathrm{C}}}
\newcommand{\caTph}{T_{\mathrm{ph}}^{\mathrm{C}}}

\newcommand{\qmH}{\hat{H}}
\newcommand{\qmA}{\hat{A}}
\newcommand{\qmrho}{\hat{\rho}}
\newcommand{\Tr}[1]{\mathrm{Tr}\left[#1\right]}
\def\<#1>{\langle #1 \rangle_q}

\def\Bzeta{\zeta_{\mathrm{B}}}
\def\Hzeta{\zeta_{\mathrm{H}}}

\newcommand{\kakunin}[1]{}

\newcommand{\soutkakunin}[1]{}

\newcommand{\memo}[1]{}

\begin{document}
\title{Thermodynamic quantities of independent harmonic oscillators in microcanonical and canonical ensembles in the Tsallis statistics}
\author{Masamichi Ishihara}
\thankstext{e1}{email: m\_isihar@koriyama-kgc.ac.jp}
\institute{Department of Food and Nutrition, Koriyama Women's University, Koriyama, Fukushima, 963-8503, JAPAN \label{addr1}}

\abstractdc{
  We study the energy and entropies for $N$ independent harmonic oscillators
  in the microcanonical and the canonical ensembles in the Tsallis classical and the Tsallis quantum statistics of entropic parameter $q$,
  where $N$ is the number of the oscillators and the value of $q$ is larger than one.
  The energy and entropies are represented with the physical temperature,
  and
  the well-known expressions are obtained for the energy and R\'enyi entropy.
  The difference between the microcanonical and the canonical ensembles is the existence of
  the condition for $N$ and $q$ in the canonical ensemble: $N(q-1)<1$.  
  The condition does not appear in the microcanonical ensemble.
  The entropies are $q$-dependent in the canonical ensemble, and are not $q$-dependent in the microcanonical ensemble.
  For $N(q-1)<1$, this difference in entropy is quite small,
  and the entropy in the canonical ensemble does not differ from the entropy in the microcanonical ensemble substantially.
  }

\maketitle    

\section{Introduction}

Power-like distributions appear in various fields of science. 
The Tsallis statistics is a possible extension of the Boltzmann-Gibbs statistics,
and the statistics gives a power-like distribution.
The Tsallis statistics has a parameter $q$, and the value $|1-q|$ is the measure of the deviation from the Boltzmann-Gibbs statistics.
This statistics has been applied to various phenomena \cite{TsallisBook}. \kakunin{gravitational force, XY model etc. きりがない}

The system composed of harmonic oscillators is basic to describe the phenomena.
The potential is often approximated with a harmonic potential, 
and a field is decomposed into harmonic oscillators in field theories.
The system of harmonic oscillators is a good base to calculate physical quantities.
Therefore, the study of independent harmonic oscillators is required in the Tsallis statistics \cite{Ishihara:2021-B},
in addition to the study of a harmonic oscillator \cite{Tsallis1998}.

It is worth to study simple models in different ensembles.
In the Boltzmann-Gibbs statistics,
the quantities are calculated in several ensembles such as microcanonical and canonical ensembles, 
and the value of a thermodynamic quantity in the microcanonical ensemble is the same as that in the canonical ensemble.
In the unconventional statistics, the equivalence between the ensembles is not always clear. 
For example, long range force may alter the properties of the statistics.

It is not always easy to calculate the quantities of harmonic oscillators in the Tsallis statistics. 
A norm equation \cite{Parvan2010} should be satisfied in the canonical ensemble in the Tsallis statistics.
The physical temperature
\cite{Kalyana:2000,Abe-PLA:2001,S.Abe:physicaA:2001,Aragao:2003,Ruthotto:2003,Toral:2003,Suyari:2006,Ishihara:phi4,Ishihara:free-field,Ishihara:Thermodyn-rel}
was introduced and has been used to describe quantities in the Tsallis statistics.
The Tsallis entropy is related to the R\'enyi entropy,
and the R\'enyi entropy is related directly to the physical temperature.

The purpose of this paper is to calculate the quantities and to clarify the differences between ensembles
in the system of independent harmonic oscillators in the Tsallis statistics with escort average \cite{TsallisBook,Tsallis1998}.
We attempt to derive the energy and the entropies
in the microcanonical and the canonical ensembles in the Tsallis classical and the Tsallis quantum statistics of the entropic parameter $q$.
The constraint between the number of the oscillators $N$ and the deviation $q-1$ in the canonical ensemble is derived for $q>1$.
The expression of the energy with the physical temperature is derived, 
and the difference between the entropy in the microcanoncal ensemble and that in the canonical ensemble is given for $N(q-1)<1$.
The difference between the ensembles in the Tsallis statistics is clarified for $N$ independent harmonic oscillators.

In this paper, the energy and entropies are calculated in the microcanonical and the canonical ensembles.
The quantities are obtained in the Tsallis classical statistics in section~\ref{sec:classical}
and in the Tsallis quantum statistics in section~\ref{sec:quantum}.
The last section is assigned for conclusions.

\section{Harmonic oscillators in the Tsallis classical statistics}
\label{sec:classical}
In this section, we calculate the quantities of $N$ independent harmonic oscillators
in the microcanonical ensemble and the canonical ensemble in the Tsallis classical statistics.
Hereafter, we use the following notations:
$A^{\mathrm{MC}}$ means the value of $A$ in the microcanonical ensemble
and $A^{\mathrm{C}}$ means the value of $A$ in the canonical ensemble.

\subsection{Microcanonical ensemble in the Tsallis classical statistics}
The entropy is defined with the number of states $W$ in the microcanonical ensemble.
We begin with calculating $W$ in the Tsallis classical statistics. 

The total energy $U$ of $N$ independent harmonic oscillators is 
\begin{subequations}
\begin{align}
  &U = \sum_{j=1}^{N} E_j,\\
  &E_j = \frac{1}{2m} p_j^2 + \frac{1}{2} m \omega^2 x_j^2, \label{def:Ej}
\end{align}
\end{subequations}
where $x_j$ and $p_j$ are the position and momentum of the harmonic oscillator numbered $j$, 
$m$ is the mass, and $\omega$ is the frequency.
As is well-known, the number of states $\Omega_0(U)$ with energy less than or equal to $U$ is given by
\begin{align}
  \Omega_0(U) = \frac{1}{\Gamma(N+1)} \left( \frac{U}{\hbar\omega} \right)^N , 
\end{align}
where $\hbar$ is the Dirac's constant and $\Gamma(x)$ is the Gamma function.
The number of states $W(U,\delta U)$ between $U$ and  $U+\delta U$ is
\begin{align}
W(U,\delta U) = \frac{1}{(N-1)!} \left( \frac{U}{\hbar\omega}\right)^{N-1} \left( \frac{\delta U}{\hbar \omega} \right) . 
\end{align}

The Tsallis entropy $\mcTsallisS$ is given \cite{TsallisBook,S.Abe:PRE:2002,Moyano:EurLett:73} by
\begin{align}
\mcTsallisS = \ln_q W = \frac{W^{1-q} -1}{1-q}, 
\label{def:cl:micro}
\end{align}
where $\ln_q x$ is the $q$-logarithm function.
The R\'enyi entropy $\mcRenyiS$ is related to the Tsallis entropy \cite{S.Abe:physicaA:2001,Vives2002}:
\begin{align}
  \mcRenyiS = \frac{1}{1-q} \ln (1+(1-q) \mcTsallisS) = \ln W . 
  \label{def:RenyiS}
\end{align}
By using the Stirling's formula, we have
\begin{align}
\mcRenyiS \sim (N-1) \left[ 1 + \ln \left( \frac{U}{(N-1) \hbar \omega} \right) \right] + \ln \left(\frac{\delta U}{\hbar \omega}\right). 
\label{SRq:classical}
\end{align}
The Tsallis entropy is also obtained by substituting $W(U,\delta U)$ into Eq.~\eqref{def:cl:micro}.

We introduce the microcanonical temperature $\mcT$ and microcanonical physical temperature $\mcTph$ by
\begin{subequations}
\begin{align}
  &\frac{1}{\mcT} = \frac{\partial \mcTsallisS}{\partial U} = W^{-q} \frac{\partial W}{\partial U}, \label{def:TME}\\
  &\frac{1}{\mcTph} = \frac{\partial \mcRenyiS}{\partial U} = \frac{1}{W} \frac{\partial W}{\partial U}. \label{def:TphME}
\end{align}
\end{subequations}
Therefore, we have the relation
\begin{align}
\frac{1}{\mcT} = W^{1-q} \frac{1}{\mcTph} \label{rel:TME:TphME}.
\end{align}

It is possible to replace $N-1$ with $N$ for large $N$, and we have
\begin{align}
\frac{1}{\mcTph} = \frac{\partial \mcRenyiS}{\partial U} \sim \frac{N}{U} 
\end{align}
Therefore, we have the equation:
\begin{align}
U \sim N\mcTph. 
\label{CL-MC:eq-of-state}
\end{align}
The energy represented with $\mcTph$ is natural.

The R\'enyi entropy, Eq.~\eqref{SRq:classical}, is represented with $\mcTph$ by using Eq.~\eqref{CL-MC:eq-of-state} for $N \gg 1$:
\begin{align}
  \mcRenyiS \sim  N \ln \left( \frac{\mcTph}{\hbar \omega} \right) + N .
  \label{eqn:Sq:approximated}
\end{align}
Equation~\eqref{def:RenyiS} is rewritten as 
\begin{align}
\mcTsallisS = \frac{1-e^{-(q-1)\mcRenyiS}}{q-1}.
\end{align}
By using Eq.~\eqref{eqn:Sq:approximated}, the Tsallis entropy in the microcanonical ensemble is given by
\begin{align}
\mcTsallisS \sim \frac{1}{q-1} \left[ 1 - \left( \frac{\hbar \omega}{e \mcTph}\right)^{N(q-1)} \right] .
  \label{classical:Tsallis:ME}
\end{align}

It is possible to obtain the relation between $\mcTph$ and $\mcT$. 
By differentiating $\mcTsallisS$ with respect to $U$, we have
\begin{align}
\frac{1}{\mcT} =e^{-(q-1)\mcRenyiS} \frac{1}{\mcTph} .
\label{relation:mcT:mcTph:mcRenyiS}
\end{align}
Substituting the expression of $\mcRenyiS$, Eq.~\eqref{eqn:Sq:approximated}, into Eq.~\eqref{relation:mcT:mcTph:mcRenyiS}, 
we have the relation between $\mcTph$ and $\mcT$:
\begin{align}
\mcTph \sim \left( \frac{\hbar\omega}{e \mcT} \right)^{\frac{N(q-1)}{1+N(q-1)}} \mcT .
\end{align}

Finally, we have the expression of $U$ with $\mcT$:
\begin{align}
U \sim \left( \frac{\hbar\omega}{e \mcT} \right)^{\frac{N(q-1)}{1+N(q-1)}} N \mcT  .
\end{align}
It is possible to expand $U$ with respect to $N(q-1)$. 
We obtain
\begin{align}
  U \sim N \mcT \left\{ 1 - N (q-1) \left[\ln \left(\frac{\mcT}{\hbar\omega}\right) + 1\right] \right\} .
\label{eqn:cl:U:mcT:highT}
\end{align}
This approximate equation is valid for $|N(q-1) \ln ((\hbar \omega)/e \mcT)| \ll 1$.

\subsection{Canonical ensemble in the Tsallis classical statistics}
In the canonical ensemble, 
we adopt the escort average and calculate the partition function $Z$.
The partition function $Z$ is given by
\begin{align}
  Z = \int \frac{dx_1 dp_1 \cdots dx_N dp_N}{h^N}
  \left[ 1 - (1-q) \frac{\cabeta}{Z^{1-q}} (E_1 + E_2 + \cdots + E_N - U) \right]^{\frac{1}{1-q}},
\end{align}
where $E_j$ is given by Eq.~\eqref{def:Ej} and $\cabeta$ is the inverse canonical temperature.
To calculate $Z$, we define $J_N^{\gamma}(x)$ by 
\begin{align}
J_N^{\gamma}(x) = \int_0^{\infty} dE_1 dE_2 \cdots dE_N \left[ (1+x U)- x (E_1+E_2+\cdots+E_N) \right]^{\gamma} .
\end{align}
The recurrence relation of $J_N^{\gamma}(x)$ is
\begin{align}
  J_N^{\gamma}(x) = \frac{1}{x (\gamma+1)} J_{N-1}^{\gamma+1}(x)
  \qquad \gamma + 1 < 0, x \neq 0 .
  \label{recurrence:J}
\end{align}
Equation \eqref{recurrence:J} is used recursively, and we have
\begin{align}
  J_N^{\gamma}(x) = \frac{1}{x^N}\frac{(1+\mu U)^{\gamma+N}}{(\gamma+1)(\gamma+2)\cdots(\gamma+N)}
  \qquad \gamma + N < 0, x \neq 0 . 
\end{align}
The partition function is rewritten as 
\begin{align}
Z = \frac{1}{(\hbar \omega)^N} J_N^{\gamma}(\mu),  
\end{align}
where $\mu$ and $\gamma$ are set as follows:
\begin{subequations}
\begin{align}
  &\mu = (1-q) \cabeta/Z^{1-q}, \label{def:a}\\ 
  &\gamma = 1/(1-q). \label{def:gamma} 
\end{align}
\label{def:a-gamma}
\end{subequations}

The energy $U$ under the escort average is given by
\begin{align}
  &U =
  \frac{\displaystyle\int \frac{dx_1 dp_1 \cdots dx_N dp_N}{h^N} (E_1 + E_2 + \cdots + E_N) (p(E))^q}
       {\displaystyle\int \frac{dx_1 dp_1 \cdots dx_N dp_N}{h^N}  (p(E))^q}, \label{CL-MC:energy}\\
  &p(E) = \frac{1}{Z} \left[ 1 - (1-q) \frac{\cabeta}{Z^{1-q}} (E_1 + E_2 + \cdots + E_N - U) \right]^{\frac{1}{1-q}}.
\end{align}
The denominator of Eq.~\eqref{CL-MC:energy}, $D_U$, is 
\begin{align}
  D_U = \frac{1}{(\hbar \omega)^N Z^q} \int_0^{\infty} dE_1 dE_2 \cdots dE_N \left[ (1+\mu U)- \mu (E_1+E_2+\cdots+E_N) \right]^{q\gamma}
  = \frac{1}{(\hbar \omega)^N Z^q} J_N^{q\gamma}(\mu) . 
\end{align}
The numerator of Eq.~\eqref{CL-MC:energy}, $N_U$, is 
\begin{align}
  N_U &= 
  \left.
  \frac{1}{(\hbar \omega)^N Z^q} \int_0^{\infty} dE_1 dE_2 \cdots dE_N (E_1 + E_2 + \cdots + E_N) \left[ (1+ x U)- x (E_1+E_2+\cdots+E_N) \right]^{q\gamma}
  \right|_{x=\mu} .
\end{align}
We use the following relation:
\begin{align}
  &(E_1+E_2+\cdots+E_N) \left[ (1+x U)- x (E_1+E_2+\cdots+E_N) \right]^{q\gamma}
  \nonumber \\
  &= U \left[ (1+x U)-x (E_1+E_2+\cdots+E_N) \right]^{q\gamma}
  - \frac{1}{(q\gamma+1)} \frac{\partial}{\partial x} \left[ (1+x U)-x (E_1+E_2+\cdots+E_N) \right]^{q\gamma+1} . 
\end{align}
Therefore, we have
\begin{align}
  N_U &=
  \frac{1}{(\hbar \omega)^N Z^q} \int_0^{\infty} dE_1 dE_2 \cdots dE_N
  \Bigg\{  U \left[ (1+x U)-x (E_1+E_2+\cdots+E_N) \right]^{q\gamma}
  \nonumber \\ & \qquad 
  - \frac{1}{(q\gamma+1)} \frac{\partial}{\partial x} \left[ (1+x U)-x (E_1+E_2+\cdots+E_N) \right]^{q\gamma+1}
  \Bigg\}
  \Bigg|_{x=\mu} \nonumber \\
  &= \frac{1}{(\hbar \omega)^N Z^q}
    \left( U J_N^{q\gamma}(\mu) - \frac{1}{(q\gamma + 1)} \left[ \frac{\partial}{\partial x} J_N^{q\gamma + 1}(x) \right]_{x = \mu} \right) . 
\end{align}
We obtain the expression of $U$ with $J_{N}^{\gamma}(\mu)$:
\begin{align}
  U = \frac{N_U}{D_U}
  = \frac{\frac{1}{(\hbar \omega)^N Z^q}
    \left( U J_N^{q\gamma}(\mu) - \frac{1}{(q\gamma + 1)} \left[ \frac{\partial}{\partial x} J_N^{q\gamma + 1}(x) \right]_{x = \mu} \right)
  }
    {\frac{1}{(\hbar \omega)^N Z^q} J_N^{q\gamma}(\mu)} . 
\end{align}
This equation is reduced to 
\begin{align}
  \frac{\partial}{\partial x} \frac{(1+x U)^{q\gamma + 1 + N}}{x^N}\Bigg|_{x=\mu}  = 0 .
\end{align}
The above equation gives
\begin{align}
  U = \frac{N}{\mu (q\gamma+1)} . 
\end{align}
With Eqs.~\eqref{def:a} and \eqref{def:gamma}, we have 
\begin{align}
  U = \frac{N}{\frac{\cabeta}{Z^{1-q}}} = N \caT Z^{1-q}.  
\end{align}
The canonical physical temperature is given by
\begin{align}
\frac{1}{\caTph} = \frac{\cabeta}{Z^{1-q}} = \frac{1}{\caT Z^{1-q}} .
\end{align}
The energy $U$ represented with $\caTph$ is
\begin{align}
U = N \caTph .
\label{CL-CA:eq-of-state}
\end{align}
Equation~\eqref{CL-CA:eq-of-state} in the canonical ensemble
is equivalent to Eq.~\eqref{CL-MC:eq-of-state} in the microcanonical ensemble when $\mcTph$ equals $\caTph$.
The number of the oscillators $N$ is restricted from the above: $N < 1/(q-1)$.

We attempt to obtain the entropies by calculating the partition function.
The partition function $Z$ with Eqs.~\eqref{def:a} and \eqref{def:gamma} is explicitly given by
\begin{align}
  Z & = \left( \frac{\caTph}{\hbar \omega} \right)^N \frac{[1+(1-q)N]^{\frac{1}{1-q}+N}}{(2-q)(3-2q)\cdots((N+1)-Nq)}
  \nonumber \\ &
  = \left( \frac{\caTph}{\hbar \omega} \right)^N \frac{[1+(1-q)N]^{\frac{1}{1-q}+N}}{\displaystyle\prod_{j=1}^N ((j+1)-jq)}, 
  \qquad \frac{1}{1-q} + N < 0 .
\end{align}
The R\'enyi entropy $\caRenyiS$ is given by
\begin{align}
  \caRenyiS = \ln Z = N \ln \left(\frac{U}{N\hbar \omega}\right) + \frac{1+(1-q)N}{1-q} \ln (1+(1-q)N) - \sum_{j=1}^N \ln(1-j(q-1)). 
\end{align}
The R\'enyi entropy in the canonical ensemble is $q$-dependent, while the entropy in the microcanonical ensemble is $q$-independent.
Expanding $\caRenyiS$ with respect to $N(q-1)$ for large $N$,
we have
\begin{align}
  \caRenyiS \sim N \ln \left( \frac{\caTph}{\hbar \omega} \right) + N +  \frac{1}{2} N(q-1).
  \label{CL:CA:Renyi}
\end{align}
By using Eq.~\eqref{CL:CA:Renyi}, 
the Tsallis entropy in the canonical ensemble is given by
\begin{align} 
  \caTsallisS \sim  \frac{1}{q-1} \left[ 1 - \left(\frac{\hbar \omega}{e \caTph}\right)^{N(q-1)} \exp\left(-\frac{1}{2N} (N(q-1))^2 \right) \right] .
  \label{classical:Tsallis:CE}
\end{align}

The R\'enyi entropy $\caRenyiS$ in the canonical ensemble resembles the entropy $\mcRenyiS$ in the microcanonical ensemble.
The difference between $\caRenyiS$, Eq.~\eqref{CL:CA:Renyi}, and $\mcRenyiS$, Eq.~\eqref{eqn:Sq:approximated} , for large $N$ is
\begin{align}
  \caRenyiS - \mcRenyiS \sim \frac{1}{2} N (q-1) ,
  \label{eqn:difference:caRS:mcRS}
\end{align}
when $\mcTph$ equals $\caTph$. 
We note that $N (q-1)$ is positive and less than one in the canonical ensemble.
Therefore, $\caRenyiS$ equals $\mcRenyiS$ substantially for $0 < N (q-1) < 1$.
From Eq.~\eqref{eqn:difference:caRS:mcRS}, $\caRenyiS$ equals $\mcRenyiS$ in the Boltzmann-Gibbs limit ($q$ approaches one).  

We also calculate the difference between $\mcTsallisS$ and $\caTsallisS$ for $\mcTph = \caTph$:
\begin{align}
  \caTsallisS - \mcTsallisS
  \sim  \frac{1}{q-1} \left(\frac{\hbar \omega}{e \caTph}\right)^{N(q-1)}  \left[1 - \exp\left(-\frac{1}{2N} (N(q-1))^2 \right) \right] .
  \label{diff:Tsallis:classical}
\end{align}
We have the relation, $0 < (N(q-1))^2 < N(q-1) <1$, for $q>1$.
We obtain the following expression for $\mcTph = \caTph$ by expanding the exponential term
in the square brackets of Eq.~\eqref{diff:Tsallis:classical}:
\begin{align}
  \caTsallisS - \mcTsallisS \sim \frac{1}{2} N(q-1) \left(\frac{\hbar \omega}{e \caTph}\right)^{N(q-1)}.
\end{align}
It is easily seen from the above expression that $\caTsallisS$ equals $\mcTsallisS$ in the Boltzmann-Gibbs limit. 

\section{Harmonic oscillators in the Tsallis quantum statistics}
\label{sec:quantum}
In this section, we calculate the quantities of $N$ independent harmonic oscillators 
in the microcanonical ensemble and the canonical ensemble in the Tsallis quantum statistics.

The quantized Hamiltonian is given by
\begin{align}
  \qmH = \sum_{j=1}^N \left( \frac{1}{2m} \hat{p}_j^2 + \frac{1}{2} m \omega^2 \hat{x}_j^2\right)
  = \sum_{j=1}^N \hbar \omega \left(\hat{n}_j + \frac{1}{2} \right),
\end{align}
where $\hat{n}_j$ is the number operator.

\subsection{Microcanonical ensemble in the Tsallis quantum statistics}
We attempt to obtain the expression of $U$ by applying the standard procedure.
The energy $U$ of $N$ independent harmonic oscillators is given by 
\begin{align}
  U(n_1,n_2,\cdots,n_N) & = \left( M + \frac{N}{2} \right) \hbar \omega, 
  \qquad M = \sum_{j=1}^N n_j, 
\end{align}
where $n_j$ is an integer which is larger than or equal to zero. 
The number of states $W(M,N)$ is 
\begin{align}
W(M,N) = \left( \begin{array}{c} M+N-1 \\ M \end{array} \right)  = \frac{(M+N-1)!}{M! (N-1)!} .
\end{align}

The R\'enyi entropy $\mcRenyiS$ and the Tsallis entropy $\mcTsallisS$ are given by $\ln W$ and $\ln_q W$ respectively.
By using the Stirling's formula, we have the approximated value of $W$: 
\begin{align}
W(M,N) \sim \frac{(M+N-1)^{M+N-1}}{M^M (N-1)^{N-1}}. 
\end{align}
The R\'enyi entropy $\mcRenyiS$ is 
\begin{align}
\mcRenyiS \sim (M+N-1)\ln(M+N-1) -M \ln M - (N-1) \ln (N-1).  
\end{align}
For fixed $N$, the inverse of the microcanonical physical temperature (Eq.~\eqref{def:TphME}) is given by 
\begin{align}
  \frac{1}{\mcTph} = \frac{\partial \mcRenyiS}{\partial U}
  = \frac{1}{\hbar \omega} \frac{\partial \mcRenyiS}{\partial M}
  = \frac{1}{\hbar \omega} \ln \left( \frac{M+N-1}{M} \right).
\label{def:qm:mcTphys}
\end{align}
For large $N$, we have
\begin{align}
\frac{M+N-1}{M} \sim \frac{U+ \frac{N \hbar \omega}{2}}{U - \frac{N \hbar \omega}{2}} . 
\label{eqn:qm:U-M-N}
\end{align}
From Eqs.~\eqref{def:qm:mcTphys} and \eqref{eqn:qm:U-M-N},
the well-known expression of $U$ is obtained:
\begin{align}
  U = N\hbar\omega \left( f_{\mathrm{B}}\left(\mcTph\right) + \frac{1}{2} \right),
\label{eqn:QM-MC:eq-of-state}
\end{align}
where $f_{\mathrm{B}}(T)$ is given by 
\begin{align}
f_{\mathrm{B}}(T) =  \frac{1}{\exp\left(\frac{\hbar\omega}{T}\right) -1} .
\label{def:fB}
\end{align}

The Tsallis entropy $\mcTsallisS$ is given by
\begin{align}
\mcTsallisS =\frac{W^{1-q}-1}{1-q} .
\end{align}
The microcanonical temperature $\mcT$ (Eq.~\eqref{def:TME}) is
\begin{align}
  \frac{1}{\mcT} = \frac{\partial \mcTsallisS}{\partial U} = W^{-q}\frac{\partial W}{\partial U}
  = \frac{1}{\hbar \omega} W^{-q} \frac{\partial W}{\partial M} . 
\end{align} 
This leads to 
\begin{align}
\frac{\hbar\omega}{\mcT} =  W^{1-q} \frac{\partial}{\partial M}  \ln W . 
\end{align}
The relation between $\mcTph$ and $W$ is already given in Eq.~\eqref{def:qm:mcTphys} with $\mcRenyiS = \ln W$: 
\begin{align}
\frac{1}{\mcTph} = \frac{1}{\hbar \omega} \frac{\partial}{\partial M} \ln W . 
\end{align}
Therefore, the relation between $\mcTph$ and $\mcT$ is 
\begin{align}
\frac{1}{\mcT} = W^{1-q} \frac{1}{\mcTph}.
\label{rel:TME:TphME2}
\end{align}
This relation was already given as Eq.~\eqref{rel:TME:TphME}.

The expression of $U$ with Eq.~\eqref{rel:TME:TphME2} is given by
\begin{align}
  U = N\hbar\omega \left( \frac{1}{\exp\left(W^{q-1} \frac{\hbar\omega}{\mcT}\right) -1} + \frac{1}{2} \right) .
  \label{U:W:mcT}
\end{align}
From Eq.~\eqref{U:W:mcT}, we have the energy $U$ represented with $\mcT$:
\begin{align}
  U \sim N\hbar\omega
  \Bigg\{
  & \left( f_{\mathrm{B}}(\mcT) + \frac{1}{2} \right)
  \nonumber \\ 
  & + N (1-q) 
  \left[ \left( f_{\mathrm{B}}(\mcT) + 1 \right) \ln \left( f_{\mathrm{B}}(\mcT) + 1 \right) -  f_{\mathrm{B}}(\mcT) \ln f_{\mathrm{B}}(\mcT) \right]
  \nonumber \\
  & \quad \times
  f_{\mathrm{B}}(\mcT)  \left( f_{\mathrm{B}}(\mcT) + 1 \right) \left( \frac{\hbar \omega}{\mcT} \right)
  \Bigg\} . 
  \label{eqn:U:mcT}
\end{align}
Equation~\eqref{eqn:U:mcT} at high $\mcT$ is 
\begin{align}
  U \sim N \mcT \Bigg\{1 - N (q-1) \left[ \ln\left(\frac{\mcT}{\hbar \omega}\right) + 1  \right]\Bigg\}, 
  \label{eqn:qm:U:mcT:highT}
\end{align}
where the condition $|N(q-1) \ln ((\hbar \omega)/e \mcT)| \ll 1$ should be satisfied in Eq.~\eqref{eqn:qm:U:mcT:highT}.
The expression of $U$ in the quantum case, Eq.~\eqref{eqn:QM-MC:eq-of-state}, coincides with
that in the classical case, Eq.~\eqref{CL-MC:eq-of-state}, at high $\mcTph$. 
The expression of $U$ in the quantum case, Eq.~\eqref{eqn:qm:U:mcT:highT}, also coincides with
that in the classical case, Eq.~\eqref{eqn:cl:U:mcT:highT}, at high $\mcT$.

The R\'enyi entropy $\mcRenyiS$ is also given by \memo{式の導出があっさりしすぎか？}
\begin{align}
  \mcRenyiS \sim
  N \left\{
  \left( 1+ f_{\mathrm{B}}\left(\mcTph\right)\right) \ln \left( 1+ f_{\mathrm{B}}\left(\mcTph\right)\right)
  -
  f_{\mathrm{B}}\left(\mcTph\right) \ln \left( f_{\mathrm{B}}\left(\mcTph\right) \right)
  \right\}.
\end{align}
At high $\mcTph$, we have
\begin{align}
  \mcRenyiS \sim
  N \ln\left( \frac{\mcTph}{\hbar \omega} \right) + N.
  \label{QM:MC:Renyi}
\end{align}
Equation \eqref{QM:MC:Renyi} is the same as Eq.~\eqref{eqn:Sq:approximated}.
Therefore, the expression of the Tsallis entropy is given by Eq.~\eqref{classical:Tsallis:ME}.

\subsection{Canonical ensemble in the Tsallis quantum statistics}
The density operator $\qmrho$ under the escort average in the canonical ensemble is given by
\begin{subequations}
\begin{align}
  & \qmrho = Z^{-1} \left( 1 - (1-q) \frac{\cabeta}{c_q} (\qmH-U) \right)^{\frac{1}{1-q}} , \\
  & c_q = \Tr{\qmrho^q}, \\
  & Z = \Tr{\left( 1 - (1-q) \frac{\cabeta}{c_q} (\qmH-U) \right)^{\frac{1}{1-q}}} , 
\end{align}
\end{subequations}
where the escort average  $\<\qmA>$ of an operator $\qmA$ is defined by 
\begin{align}
\<\qmA> = \frac{\Tr{\qmrho^q \qmA}}{\Tr{\qmrho^q}} .  
\end{align}
The following relation between $c_q$ and $Z$ is used to find the expressions of physical quantities:
\begin{align}
c_q = Z^{1-q} .
\label{rel:cq-Z}
\end{align}

The partition function $Z$ is calculated as 
\begin{subequations}
\begin{align}
  Z = \sum_{n_1,\cdots,n_N=0}^{\infty} \left( ab  + a (n_1+\cdots+n_N) \right)^{\frac{1}{1-q}} 
    = a^{\frac{1}{1-q}} \sum_{n_1,\cdots,n_N=0}^{\infty} \frac{1}{\left( b  + n_1+n_2+\cdots+n_N \right)^{\frac{1}{q-1}}} . 
\end{align}
where
\begin{align}
  &a =  (q-1) \left(\frac{\cabeta \hbar \omega}{c_q}\right), \\
  &ab = 1+ (q-1) \left( \frac{\cabeta}{c_q} \right) \left(\frac{1}{2} N\hbar\omega - U \right). \label{eq:ab}
\end{align}  
\end{subequations}
The partition function $Z$ is represented with the Barnes zeta function $\Bzeta$:
\begin{align}
Z =  a^{\frac{1}{1-q}} \Bzeta\left(\frac{1}{q-1}, b; N\right), 
\label{QM-CA:Z:Bzeta}
\end{align}
where $\Bzeta$ (see also appendix~\ref{sec:zeta}) is given by
\begin{align}
\Bzeta (s, \alpha; N) = \sum_{n_1,\cdots,n_N=0}^{\infty} \frac{1}{ ( \alpha + n_1 + n_2 + \cdots + n_N)^s} .
\end{align}
In the same way, $c_q$ is calculated directly by using the definition $c_q=\Tr{\qmrho^q}$:
\begin{align}
c_q = \frac{1}{Z^q} a^{\frac{q}{1-q}} \Bzeta\left(\frac{q}{q-1}, b; N\right) .
\end{align}
We obtain the following norm equation from Eq.~\eqref{rel:cq-Z}: 
\begin{align}
a \Bzeta \left(\frac{1}{q-1}, b; N \right) = \Bzeta\left(\frac{q}{q-1}, b; N\right) .
\label{eqn:QM-CL:self-consistent}
\end{align}

We obtain the energy $U$ approximately from the norm equation, Eq.~\eqref{eqn:QM-CL:self-consistent}.
For large $\alpha$, we have
\begin{align}
  \Bzeta(1+z, \alpha; N)
  \sim \frac{1}{\displaystyle\prod_{j=0}^{N-1} (z-j)} \frac{1}{\alpha^{z-(N-1)}}
  + O\left(\frac{1}{\alpha^{1+z-(N-1)}}\right)
    \qquad (1+z > N) . 
\label{Bzeta:appro}
\end{align}
Applying Eq.~\eqref{Bzeta:appro} to Eq.~\eqref{eqn:QM-CL:self-consistent},
we obtain the product $ab$ approximately for large $b$ (high physical temperature):
\begin{align}
  a b
  = \frac{\displaystyle\prod_{j=0}^{N-1} ((2-q)-j(q-1))}{\displaystyle\prod_{j=0}^{N-1} (1-j(q-1))}
  = \frac{\displaystyle\prod_{j=0}^{N-1} (1-(j+1)(q-1))}{\displaystyle\prod_{j=0}^{N-1} (1-j(q-1))}  
  = 1 - N (q-1)  \qquad N < (q-1)^{-1} . 
\end{align}
Equation \eqref{eq:ab} gives the energy $U$:
\begin{align}
 U = N \caTph + \frac{1}{2} N \hbar \omega . 
\label{QM-Canonical:U}
\end{align}

We also calculate the escort average of the Hamiltonian directly: 
\begin{align}
  U = \frac{\Tr{\qmrho^q \qmH}}{\Tr{\qmrho^q}} 
  =  \frac{1}{2} N \hbar \omega 
  + \frac{\hbar \omega a^{\frac{q}{1-q}}}{Z} \left(\Bzeta \left(\frac{1}{q-1}, b ;N\right) - b \Bzeta \left(\frac{q}{q-1}, b ;N\right)  \right) .
\label{expression:U:QMcannoical:basic}
\end{align}
Another form of the energy $U$ is also derived: 
\begin{align}
U =\frac{\hbar\omega}{Z} a^{\frac{q}{1-q}} 
\left(\Bzeta \left(\frac{q}{q-1}, b ;N\right) + \left(\frac{N}{2}  - b\right)\Bzeta \left(\frac{1}{q-1}, b ;N\right) \right) .
\label{expression:U:QMcanonical}
\end{align}
It is possible to calculate Eq.~\eqref{expression:U:QMcannoical:basic} approximately with Eq.~\eqref{QM-CA:Z:Bzeta}
by using the approximate expression of Barnes zeta function, Eq.~\eqref{Bzeta:appro}.
For large $b$, we obtain
\begin{align}
  \frac{\Bzeta \left(\frac{q}{q-1}, b ;N\right)}{\Bzeta \left(\frac{1}{q-1}, b ;N\right)} \sim b^{-1} [ 1 - N (q-1)]
  \qquad N < (q-1)^{-1}.
\end{align}
We have Eq.~\eqref{QM-Canonical:U} again from Eq.~\eqref{expression:U:QMcannoical:basic}.

The R\'enyi entropy $\caRenyiS$ can be calculated with the partition function $Z$: $\caRenyiS = \ln Z$.
For large $b$, the partition function $Z$ is approximated as
\begin{align}
  Z = a^{\frac{1}{1-q}} \zeta_B \left( \frac{1}{q-1}, b; N \right)
  \sim \frac{(q-1)^N}{\displaystyle\prod_{j=0}^{N-1} (1-(j+1)(q-1))}  \frac{b^N}{(ab)^{1/(q-1)-N} } . 
\end{align}
We have the following expression of $\caRenyiS$ for large $N$ by expanding $\ln Z$ with respect to $N(q-1)$: 
\begin{align}
  \caRenyiS \sim N \ln \left( \frac{\caTph}{\hbar \omega} \right) + N + \frac{1}{2}  N (q-1).
  \label{QM:CA:Renyi}
\end{align}
Equation \eqref{QM:CA:Renyi} is the same as Eq.~\eqref{CL:CA:Renyi}.
Therefore, the expression of the Tsallis entropy is given by Eq.~\eqref{classical:Tsallis:CE}.

We obtain the difference between $\caRenyiS$, Eq.~\eqref{QM:CA:Renyi}, and $\mcRenyiS$, Eq.~\eqref{QM:MC:Renyi},
at high physical temperature in the quantum statistics.
The difference is 
\begin{align}
\caRenyiS - \mcRenyiS \sim \frac{1}{2} N (q-1), 
\end{align}
when $\caTph$ equals $\mcTph$.
We note that the quantity $N(q-1)$ is restricted below one in the canonical ensemble.
Therefore, $\caRenyiS$ equals $\mcRenyiS$ substantially for $0 < N (q-1) < 1$, as shown in the classical statistics.
These results in the quantum statistics coincide with those in the classical statistics.
At high physical temperature,
the difference between $\caTsallisS$ and $\mcTsallisS$ in quantum statistics is the same as the difference in the classical statistics.
Therefore, the difference is given by Eq.~\eqref{diff:Tsallis:classical}.

\section{Conclusions}

We studied the thermodynamic quantities in the Tsallis classical and the Tsallis quantum statistics of the entropic parameter $q$,
where the value of $q$ is larger than one. 
We treated the $N$ independent harmonic oscillators with the same frequencies,
where $N$ is the number of the oscillators.
The energy was represented with the physical temperature,
and the R\'enyi entropy in the microcanonical ensemble was compared with the R\'enyi entropy in the canonical ensemble.
The Tsallis entropy was calculated with the relation between the Tsallis entropy $S_{\mathrm{T}q}$ and the R\'enyi entropy $S_{\mathrm{R}q}$:
$S_{\mathrm{R}q}=(1-q)^{-1} \ln (1+(1-q) S_{\mathrm{T}q})$,
and the Tsallis entropy in the microcanonical ensemble was compared with the Tsallis entropy in the canonical ensemble.

The physical temperature $\Tph$ is less than or equal to the temperature $T$ for $q>1$.
The microcanonical physical temperature $\mcTph$ and the microcanonical temperature $\mcT$ has the relation $\mcTph = W^{1-q} \mcT$,
where $W$ is the number of states.
The quantity $W^{q-1}$ is larger than or equal to one for $W \ge 1$ and $q>1$.
Therefore, the physical temperature $\mcTph$ is less than or equal to the temperature $\mcT$.
The canonical physical temperature $\caTph$ and the canonical temperature $\caT$ has the relation $\caTph = Z^{1-q} \caT$,
where $Z$ is the partition function.
The partition function $Z$ is larger than or equal to one for $\caRenyiS \ge 0$ 
because of the relation $Z=\exp(\caRenyiS)$,
where $\caRenyiS$ is the R\'enyi entropy in the canonical ensemble.
The canonical physical temperature $\caTph$ is less than or equal to the temperature $\caT$ for $q>1$.

The condition $N(q-1)<1$ appears in the canonical ensemble, while this condition does not appear in the microcanonical ensemble.
Such conditions were already obtained in the previous studies \cite{Ishihara:2021-B,Abe-PLA:2001}.
As expected, the maximum value of $N$ which satisfies $N(q-1)<1$ goes to infinity, as $q$ approaches one.
This constraint is explained by the power-law behavior of the distribution for $q>1$ in the Tsallis statistics.
This condition does not appear for $q=1$ (the Boltzmann-Gibbs statistics), because the distribution decreases exponentially.
The existence of the restriction is the apparent difference between the microcanonical ensemble and the canonical ensemble
in the Tsallis statistics.

It was shown that the energy $U$ represented with the physical temperature $\Tph$ is not $q$-dependent at high $\Tph$ in both the ensembles:
the energy has the well-known relation $U = N \Tph$ at high physical temperature.
At high physical temperature, the expression of $U$ represented with $\Tph$ in the Tsallis statistics is the same as that in the Boltzmann-Gibbs statistics.

The $q$-dependence of the R\'enyi entropy in the canonical ensemble slightly differs from that in the microcanonical ensemble.
The R\'enyi entropy represented with the physical temperature is not $q$-dependent in the microcanonical ensemble.
Therefore, the Tsallis entropy with the physical temperature is not $q$-dependent in the microcanonical ensemble.
This result is consistent with the result within Tsallis formalism in microcanonical ensemble \cite{Toral:2003}.
In contrast, the R\'enyi entropy has the $q$-dependent term, $N(q-1)/2$, in the canonical ensemble. 
The R\'enyi entropy in the canonical ensemble is not different from the R\'enyi entropy in the microcanonical ensemble substantially,
because $N (q-1)$ is less than one.
In the same way, 
the Tsallis entropy in the canonical ensemble is not different from the Tsallis entropy in the microcanonical ensemble substantially,
because $N(q-1)$ is less than one.

In this paper,
we discussed the difference between the quantity in the microcanonical ensemble and the quantity in the canonical ensemble
in the Tsallis statistics of the entropic parameter $q$ for $N$ independent harmonic oscillators.
The restriction $N<(q-1)^{-1}$ appears in the canonical ensemble.
The various properties of the unconventional statistics will be studied in the future.

\medskip\noindent\textbf{Data availability statement}
This manuscript has no associated data or the data will not be deposited.
[Authors' comment: This study is theoretical, and no data is generated.]

\appendix
\section{Hurwitz zeta function and Barnes zeta function}
\label{sec:zeta}
In this appendix, we give the approximate expressions of Hurwitz and Barnes zeta functions.
The derivation is also given in the appendices A and B of the reference \cite{Ishihara:2021-B} .

The Hurwitz zeta function $\Hzeta$ is defined by
\begin{equation}
\Hzeta(s, \alpha) = \sum_{n=0}^{\infty} \frac{1}{(\alpha+n)^s}.
\end{equation}
In this appendix, we treat the case of $s>1$ and $\alpha>0$.

Applying the Euler-Maclaurin formula,
we have 
\begin{align}
  \Hzeta(1+z, \alpha)
  &= \frac{1}{z\alpha^z} + \frac{1}{2\alpha^{1+z}} + \sum_{k=1}^{M-1} \frac{(-1)^{k+1} B_{k+1}}{(k+1)!} \frac{\Gamma(z+k+1)}{\Gamma(z+1)} \frac{1}{\alpha^{z+k+1}}
  \nonumber \\
  & \quad - \frac{(-1)^M}{M!} \int_0^{\infty} dx B_M(x-[x]) f^{(M)}(x) \qquad (z>0, \alpha > 0) ,
\label{eqn:Hzeta}
\end{align}
where $B_k$ is Bernoulli number.
The Hurwitz zeta function can be expressed in the other forms  \cite{Boumali2014}.
From Eq.~\eqref{eqn:Hzeta}, we find that the $\Hzeta(1+z, \alpha)$ for $\alpha \gg 1$ behaves as 
\begin{align}
  \Hzeta(1+z,\alpha) \sim \frac{1}{z \alpha^{z}}.
  \label{eqn:Hzeta:large}
\end{align}

The Barnes zeta function \cite{Ruijsenaars:2000,Kirsten:2010} is defined by
\begin{align}
  & \Bzeta(s,\alpha|\vec{\omega}_N) = \sum_{n_1,\cdots,n_N=0}^{\infty} \frac{1}{(\alpha + \omega_1 n_1 + \cdots + \omega_N n_N)^s}
  \qquad \vec{\omega}_N = (\omega_1, \omega_2, \cdots, \omega_N), 
  \label{def:Barnes-zeta}
\end{align}
where $s>N$, $\alpha > 0$, and $\omega_j > 0$.
The Barnes zeta function for sufficiently large $\alpha$ has the following relation
\begin{align}
  \Bzeta(1+z,\alpha|\vec{\omega}_N) 
  \sim \frac{1}{z \omega_N} \Bzeta(z, \alpha|\vec{\omega}_{N-1}) .
\label{eq:recurrence-relation}
\end{align}
This relation is derived with Eq.~\eqref{eqn:Hzeta:large}.
By using the recurrence relation, Eq.~\eqref{eq:recurrence-relation},
we have the following approximate expression of $\Bzeta$ for $\alpha \gg 1$:
\begin{align}
  \Bzeta(1+z, \alpha|\vec{\omega}_N)
  &\sim
  \frac{1}{\Bigg(\displaystyle\prod_{j=0}^{N-1} (z-j)\Bigg) \Bigg(\displaystyle\prod_{j=1}^{N} \omega_j \Bigg) \alpha^{z-(N-1)}} 
  \qquad\qquad (z - (N-1) > 0). 
  \label{eqn:Bzeta:approximation}
\end{align}

In the present case, $\vec{\omega}_N$ is set to $\vec{\omega}_N = (1,1,\cdots,1)$.
For simplicity, we use the following notation for the Barns zeta function $\Bzeta$ with $\vec{\omega}_N = (1,1,\cdots,1)$:
\begin{align}
\Bzeta (s, \alpha; N) = \sum_{n_1,\cdots,n_N=0}^{\infty} \frac{1}{ ( \alpha + n_1 + n_2 + \cdots + n_N)^s} .
\label{Bzeta:simple}
\end{align}
The approximate expression of $\Bzeta (1+z, \alpha; N)$ for $\alpha \gg 1$ is 
\begin{align}
  \Bzeta (1+z, \alpha; N) \sim \frac{1}{\Bigg(\displaystyle\prod_{j=0}^{N-1} (z-j)\Bigg) \alpha^{z-(N-1)}} 
  \qquad\qquad (z - (N-1) > 0). 
\end{align}


\end{document}